\documentclass[twocolumn]{aastex63}

\usepackage{longtable}
\usepackage{graphicx}
\usepackage{txfonts}
\usepackage{verbatim}
\usepackage{color}
\usepackage{threeparttablex}
\submitjournal{ApJ}

\usepackage{natbib}
\defcitealias{DeSomma2020}{DS2020}
\defcitealias{Ripepi2019}{Ripepi2019}
\defcitealias{Bhardwaj2017}{Bhardwaj2017}

\submitjournal{ApJ}
\shorttitle{RR Lyrae PLZ relations in the Rubin-LSST filters}
\shortauthors{Marconi et al.}

\graphicspath{{./}{figures/}}

\begin{document}

\title{New theoretical Period-Luminosity-Metallicity relations for RR Lyrae in the Rubin-LSST filters}

\correspondingauthor{Marcella Marconi}
\email{marcella.marconi@inaf.it}

\author{Marcella Marconi}
\affiliation{ INAF-Osservatorio astronomico di Capodimonte \\
Via Moiariello 16 \\
80131 Napoli, Italy}

\author{Roberto Molinaro}
\affiliation{ INAF-Osservatorio astronomico di Capodimonte \\
Via Moiariello 16 \\
80131 Napoli, Italy}

\author{Massimo Dall'Ora}
\affiliation{ INAF-Osservatorio astronomico di Capodimonte \\
Via Moiariello 16 \\
80131 Napoli, Italy}

\author{Vincenzo Ripepi}
\affiliation{ INAF-Osservatorio astronomico di Capodimonte \\
Via Moiariello 16 \\
80131 Napoli, Italy}

\author{Ilaria Musella}
\affiliation{ INAF-Osservatorio astronomico di Capodimonte \\
Via Moiariello 16 \\
80131 Napoli, Italy}

\author{Giuseppe Bono}
\affiliation{Università degli Studi di Roma "Tor Vergata" \\
Via Cracovia, 50\\
00133 Roma, Italy}

\author{Vittorio Braga}
\affiliation{Instituto de Astrofísica de Canarias \\
La Laguna\\
E-38205, Spain}

\author{Marcella Di Criscienzo}
\affiliation{INAF – Osservatorio Astronomico di Roma \\
Via Frascati 33\\
00078 Monte Porzio Catone (RM), Italy}

\author{Giuliana Fiorentino}
\affiliation{INAF – Osservatorio Astronomico di Roma \\
Via Frascati 33\\
00078 Monte Porzio Catone (RM), Italy}

\author{Silvio Leccia}
\affiliation{ INAF-Osservatorio astronomico di Capodimonte \\
Via Moiariello 16 \\
80131 Napoli, Italy}

\author{Matteo Monelli}
\affiliation{Instituto de Astrofísuca de Canarias \\
La Laguna\\
E-38205, Spain}

\begin{abstract}
\noindent 
The revolutionary power of future Rubin-LSST observations will allow us to significantly improve the physics of pulsating stars, including RR Lyrae. In this context, an updated theoretical scenario predicting all the relevant pulsation observables in the corresponding photometric filters is mandatory. The bolometric light curves based on a recently computed extensive set of nonlinear convective pulsation models for RR Lyrae stars, covering a broad range in metal content and transformed into the Rubin-LSST photometric system. Predicted Rubin-LSST mean magnitudes and pulsation amplitudes have been adopted to built the Bailey diagrams (luminosity amplitude vs period) and the
color-color diagrams in these bands. The current findings indicate that the $g_{LSST}-r_{LSST}$, $r_{LSST}-i_{LSST}$ colors obey to a well defined linear relation with the metal content. Moreover, the Period
Luminosity relations display in the reddest filters ($r_{LSST},i_{LSST},z_{LSST},y_{LSST}$) a
signiﬁcant dependence on the assumed metal abundance. In particular, more metal-rich RR Lyrae are predicted to be fainter at fixed period. Metal-dependent  Period-Wesenheit relations for different combinations of optical and NIR filters are also provided. These represent powerful tools to infer individual distances independently of reddening uncertainties, once the metal abundance is known and no relevant deviations from the adopted extinction law occur.
Finally, we also derived new linear and quadratic absolute magnitude
metallicity relations ($g_{LSST}$ vs $[Fe/H]$) and the metallicity coeﬃcient
is consistent with previous findings concerning the $B$ and the $V$ band.

\end{abstract}


\section{Introduction}
RR Lyrae stars are well known pulsating variables, currently adopted as standard candles and tracers of intrinsic properties in old stellar populations. Their role as distance indicators is largely based on the existence of a  Luminosity-metallicity relation in the optical bands, namely Johnson B and V \citep[see][and references therein]{Caputo2000,2004ApJ...612.1092D,2018ApJ...864L..13M}, and a  Period-Luminosity (PL) relation in the Near-Infrared (NIR) filters \citep[see e.g.][and references therein]{Longmore86,Longmore90,2006MmSAI..77..214D,2011MNRAS.416.1056C,2015ApJ...808...50M,2015ApJ...807..127M}. In spite of the well known physical basis \citep[see e.g.]{2001MNRAS.326.1183B,2004ApJS..154..633C}, the power of the  NIR PL relation 
has not been fully exploited yet, due to the quite debated dependence on metallicity. 
Indeed, nonlinear convective pulsation models of RR Lyrae at different metallicities predict a non negligible metal dependence \citep[see e.g.][and references therein]{2003MNRAS.344.1097B,2015ApJ...808...50M}. On the other hand, empirical studies \citep[e.g.][]{2006MNRAS.372.1675S,2006MmSAI..77..214D} provided smaller metallicity effects, not always consistent with each other. More recently, several
determinations \citep[see e.g.][]{2017ApJ...838..107S,2018MNRAS.481.1195M}
seem to be more in agreement with the predicted coefficient by
\citet{2015ApJ...808...50M}, 
with values in the range 0.16-0.18 mag $dex^{-1}$ for the coefficient of the [Fe/H] term in the K band linear Period-Magnitude-[Fe/H] (PLZ) relation.
\citet{2021MNRAS.500.5009M} relied on the extended metal-dependent model set presented in \citet{2015ApJ...808...50M} to  derive the  theoretical light curves in the Gaia bands $G$, $G_{BP}$, and $G_{RP}$ and, in turn, the  intensity-weighted mean magnitudes and pulsation amplitudes in these filters. The former  were used to obtain the first theoretical Period-Wesenheit relations for RR Lyrae in the Gaia filters \citep[see][for details]{2021MNRAS.500.5009M}, whereas the latter were combined with the pulsation periods to investigate the behaviour of the Bailey (e.g. period-amplitude) diagrams in the Gaia filters. The inferred theoretical Period-Wesenheit (PW) relations were applied to Galactic RR Lyrae in the Gaia Data Release 2 database with individual  metal abundances to obtain theoretical individual parallaxes that were found to be in very good agreement with Gaia astrometric results.
The next Gaia Data Release (DR3) will be in the summer of 2022, but it  was anticipated by an Early DR3 (EDR3) in December 2020.  
The increasing accuracy of Gaia results will allow us to better and better constrain the individual distances of Galactic RR Lyrae and at the same time the physical and numerical assumptions in model computations.
The extension of these capabilities to further distances (up to 5 mag fainter) will be possible thanks to the upcoming revolutionary Vera C. Rubin Observatory Legacy Survey of Space and Time (Rubin-LSST). 
The Rubin-LSST will image the southern night sky repeatedly in the $u,\,g,\,r,\,i,\,z,\, y$ filters, at limiting magnitudes fainter than 25 mag and excellent image quality, thus allowing us to obtain extremely well sampled multi-band light curves of various classes of pulsating stars, including RR Lyrae stars.
In this context, it is mandatory to provide a theoretical scenario in the Rubin-LSST filters, with the capability of predicting all the relevant pulsation observables, to pave the way to productive comparisons between theory and observations. 
In this paper we present some of these pulsation properties as predicted by nonlinear convective hydrodynamic models at different 
chemical compositions \citep[see][for details]{2015ApJ...808...50M} in the Rubin-LSST filters. In Section \ref{sec-models} we present the atlas of predicted  light curves in the LSST filters, for both F-mode and FO-mode pulsators and the corresponding metal-dependent Bailey diagrams. In Section \ref{sec-colcol} the predicted behaviour in the various color-color planes is investigated. In Section \ref{sec-pl-pw} we derive the first theoretical metal dependent PL relations in the $r_{LSST}$, $i_{LSST}$, $z_{LSST}$ and $y_{LSST}$ filters as well as PW relations for various optical and NIR filter combinations. In Section \ref{sec-g-feh} we present the first theoretical $g_{LSST}$ magnitude versus $[Fe/H]$ relation both in the traditionally linear and quadratic form over the whole adopted metallicity range. Some final remarks close the paper.

\begin{table}
\centering
\tabcolsep=0.11cm
\begin{tabular}{lcccc}
  \hline
$Z$ & $Y$ & $M$ & $\log(L/L_\odot)$ &  \\ 
 &  & $(M_\odot)$ & $(dex)$ &\\
  \hline
0.0001 & 0.245 & 0.800 & 1.76 & ZAHB \\ 
  0.0001 & 0.245 & 0.800 & 1.86 & Brighter \\ 
  0.0001 & 0.245 & 0.720 & 1.96 & Evolved \\ 
  0.0003 & 0.245 & 0.716 & 1.72 & ZAHB \\ 
  0.0003 & 0.245 & 0.716 & 1.82 & Brighter \\ 
  0.0003 & 0.245 & 0.650 & 1.92 & Evolved \\ 
  0.0006 & 0.245 & 0.670 & 1.69 & ZAHB \\ 
  0.0006 & 0.245 & 0.670 & 1.79 & Brighter \\ 
  0.0006 & 0.245 & 0.600 & 1.89 & Evolved \\ 
  0.0010 & 0.245 & 0.640 & 1.67 & ZAHB \\ 
  0.0010 & 0.245 & 0.640 & 1.77 & Brighter \\ 
  0.0010 & 0.245 & 0.580 & 1.87 & Evolved \\ 
  0.0040 & 0.250 & 0.590 & 1.61 & ZAHB \\ 
  0.0040 & 0.250 & 0.590 & 1.71 & Brighter \\ 
  0.0040 & 0.250 & 0.530 & 1.81 & Evolved \\ 
  0.0080 & 0.256 & 0.570 & 1.58 & ZAHB \\ 
  0.0080 & 0.256 & 0.570 & 1.68 & Brighter \\ 
  0.0080 & 0.256 & 0.510 & 1.78 & Evolved \\ 
  0.0200 & 0.270 & 0.540 & 1.49 & ZAHB \\ 
  0.0200 & 0.270 & 0.540 & 1.59 & Brighter \\ 
  0.0200 & 0.270 & 0.510 & 1.69 & Evolved \\ 
   \hline
\end{tabular}
\caption{This table contains the elemental abundance Z and the Helium abundance Y of the considered models, respectively
in the column 1 and 2, the mass M and the luminosity $\log L/L_\odot$ respectively in the columns 3 and 4, while in the column 5 a 
string tag is contained labeling the ZAHB, the Brighter and the Evolved models introduced in the text.} 
\label{tab-main-ZYML-params}
\end{table}

\begin{table*}
\centering
\tabcolsep=0.11cm
\begin{tabular}{cccccccccccccccccc}
  \hline
Z & Y & P & M & $\log(L/L_\odot)$ & $T_{eff}$ & $ \langle u_{LSST} \rangle$ & A$_u$ & $ \langle g_{LSST} \rangle$ & A$_g$ & $ \langle r_{LSST} \rangle$ & A$_r$ & $ \langle i_{LSST} \rangle$ & A$_i$ & $ \langle z_{LSST} \rangle$ & A$_z$ & $ \langle y_{LSST} \rangle$ & A$_y$ \\  
 &  & (days) & ($M_\odot$) & (dex) & (K) & (mag) & (mag) & (mag) & (mag) & (mag) & (mag) & (mag) & (mag) & (mag) & (mag) & (mag) & (mag) \\  
  \hline
0.0001 & 0.245 & 0.7624 & 0.80 & 1.760 & 6000 & 1.375 & 0.335 & 0.461 & 0.379 & 0.245 & 0.274 & 0.187 & 0.212 & 0.189 & 0.179 & 0.191 & 0.173 \\ 
  0.0001 & 0.245 & 0.7210 & 0.80 & 1.760 & 6100 & 1.366 & 0.654 & 0.445 & 0.752 & 0.245 & 0.559 & 0.196 & 0.438 & 0.203 & 0.375 & 0.205 & 0.369 \\ 
  0.0001 & 0.245 & 0.6839 & 0.80 & 1.760 & 6200 & 1.354 & 0.833 & 0.426 & 0.958 & 0.244 & 0.725 & 0.206 & 0.575 & 0.219 & 0.502 & 0.222 & 0.501 \\ 
  0.0001 & 0.245 & 0.6469 & 0.80 & 1.760 & 6300 & 1.343 & 0.879 & 0.407 & 1.013 & 0.245 & 0.765 & 0.219 & 0.607 & 0.238 & 0.531 & 0.241 & 0.532 \\ 
  0.0001 & 0.245 & 0.6133 & 0.80 & 1.760 & 6400 & 1.330 & 0.932 & 0.385 & 1.067 & 0.247 & 0.822 & 0.234 & 0.669 & 0.259 & 0.597 & 0.263 & 0.600 \\ 
  0.0001 & 0.245 & 0.5810 & 0.80 & 1.760 & 6500 & 1.315 & 1.184 & 0.362 & 1.336 & 0.250 & 1.002 & 0.253 & 0.797 & 0.284 & 0.710 & 0.288 & 0.718 \\ 
   \hline
\end{tabular}
\caption{Fundamental model parameters: the values of Z and Y are listed in columns 1 and 2, respectively, 
the period of pulsation is in column 3, the mass, luminosity and effective temperatures are listed 
in columns 4, 5, and 6, while the LSST light curve intensity average magnitudes and amplitudes are reported in the columns 
from 7 to 18 for all the selected bands. The complete table is available in electronic format.} 
\label{tab-mainParameters-F}
\end{table*}

\begin{table*}
\centering
\tabcolsep=0.11cm
\begin{tabular}{cccccccccccccccccc}
  \hline
Z & Y & P & M & $\log(L/L_\odot)$ & $T_{eff}$ & $ \langle u_{LSST} \rangle$ & A$_u$ & $ \langle g_{LSST} \rangle$ & A$_g$ & $ \langle r_{LSST} \rangle$ & A$_r$ & $ \langle i_{LSST} \rangle$ & A$_i$ & $ \langle z_{LSST} \rangle$ & A$_z$ & $ \langle y_{LSST} \rangle$ & A$_y$ \\  
 &  & (days) & ($M_\odot$) & (dex) & (K) & (mag) & (mag) & (mag) & (mag) & (mag) & (mag) & (mag) & (mag) & (mag) & (mag) & (mag) & (mag) \\  
  \hline
0.0001 & 0.245 & 0.4107 & 0.80 & 1.760 & 6600 & 1.301 & 0.607 & 0.346 & 0.710 & 0.243 & 0.528 & 0.255 & 0.411 & 0.297 & 0.354 & 0.304 & 0.356 \\ 
  0.0001 & 0.245 & 0.3910 & 0.80 & 1.760 & 6700 & 1.292 & 0.756 & 0.327 & 0.873 & 0.246 & 0.665 & 0.272 & 0.532 & 0.319 & 0.472 & 0.325 & 0.477 \\ 
  0.0001 & 0.245 & 0.3721 & 0.80 & 1.760 & 6800 & 1.282 & 0.962 & 0.310 & 1.090 & 0.254 & 0.805 & 0.293 & 0.630 & 0.345 & 0.558 & 0.351 & 0.568 \\ 
  0.0001 & 0.245 & 0.3553 & 0.80 & 1.760 & 6900 & 1.272 & 1.111 & 0.297 & 1.214 & 0.264 & 0.885 & 0.315 & 0.685 & 0.373 & 0.607 & 0.379 & 0.618 \\ 
  0.0001 & 0.245 & 0.3386 & 0.80 & 1.760 & 7000 & 1.263 & 1.133 & 0.286 & 1.212 & 0.273 & 0.880 & 0.337 & 0.680 & 0.400 & 0.605 & 0.406 & 0.615 \\ 
  0.0001 & 0.245 & 0.3235 & 0.80 & 1.760 & 7100 & 1.255 & 1.096 & 0.274 & 1.162 & 0.280 & 0.834 & 0.355 & 0.639 & 0.425 & 0.570 & 0.430 & 0.580 \\ 
   \hline
\end{tabular}
\caption{First overtone model parameters: the values of Z and Y are listed in columns 1 and 2, respectively, 
the period of pulsation is in column 3, the mass, luminosity and effective temperatures are listed 
in columns 4, 5, and 6, while the LSST light curve intensity average magnitudes and amplitudes are reported in the columns 
from 7 to 18 for all the selected bands. The complete table is available in electronic format.} 
\label{tab-mainParameters-FO}
\end{table*}

\section{Theoretical light curves in the Rubin-LSST filters}\label{sec-models}

We took into account the model set presented in \citet{2015ApJ...808...50M}, that covers a wide range of metal abundances, namely $Z=0.0001,0.0003,0.0006,0.001,0.004,0.008,0.02$. We considered two different stellar masses and three luminosity levels for each selected chemical composition, as detailed in Table ~\ref{tab-main-ZYML-params}. For each chemical composition, the higher stellar mass and the lowest luminosity level correspond to the predicted Zero Age Horizontal Branch (ZAHB) values \citep[see e.g.][and references therein]{2013AA}. To account for possible uncertainties in the evolutionary luminosity prediction  \citep[see e.g.][and references therein]{Valle2013,Cassisi2021}, a slightly brighter (by 0.1 dex) luminosity level (than the ZAHB one) for the ZAHB stellar mass is also considered, while lower stellar mass (by about $10 \%$) and still brighter models (0.2 dex more than the ZAHB level) were also computed for each $Z$ and $Y$ combination, in order to model possible evolved RR Lyrae \citep[see][and references therein, for details]{2015ApJ...808...50M}.
In particular, all the bolometric light curves of models with effective temperature within the F or FO instability strip for the tabulated chemical compositions and masses, have been transformed into the Rubin-LSST filters $u_{LSST}$, $g_{LSST}$, $r_{LSST}$, $i_{LSST}$, $z_{LSST}$ and $y_{LSST}$  bands. To perform these transformations, we used the bolometric corrections (BC) tables provided by \citet{che19}, including a wide variety of photometric systems and based on the PHOENIX spectral libraries. 
The full details of our adopted procedure are given in \citet{2021MNRAS.500.5009M}. Here we just recall that we used a proprietary C code to interpolate the \citet{che19} tables to obtain the BCs corresponding to the  log(g), $T_{eff}$ and Z values of our models. Moreover, we note that the BC tables provided by \citet{che19} for the LSST filters are in the AB system, therefore all our results hereafter are specific for this photometric system. The obtained multi-filter atlas of luminosity variations is available upon request. In Figure 1 and 2 we plot a subset of predicted light curves  converted into the selected bands, for the F and FO mode, respectively. In both figures the left panels display relatively hot models, close to the predicted blue boundary of the instability strip, the central panels in each row correspond to models located in the middle of the instability strip, whereas the right panels show model light curves close to the red boundary. The metallicity increases from $Z=0.0001$ (top panels) to $Z=0.02$ (bottom panels), with the middle panels in each column corresponding to $Z=0.001$.
Inspection of these plots confirms that the Rubin-LSST light curves are expected to show a decrease in the pulsation amplitudes as the filter wavelength increases. Moreover, independently of the adopted chemical composition, higher amplitudes are predicted for F models at the blue edge of the instability strip, with a linear decrease of F-amplitudes as the logarithm of the pulsation period increases. This linear trend is the typical feature of F-mode RR Lyrae in the Bailey diagram \citep[see Figure 3, see also][and references therein]{2004ApJ...612.1092D,2021MNRAS.500.5009M}. On the other hand, the FO-mode light curves in Figure 2 show the typical maximum of the pulsation amplitude in the central region of the instability strip and a mild decrease towards the two boundaries. 
The predicted Bailey diagrams for the ZAHB and the evolved models are shown in Figures \ref{fig-bailey} and \ref{fig-bailey-evolved}, respectively, for both F and FO pulsators, in all the Rubin-LSST filters. As noticed above, the predicted F-mode amplitudes show a systematic decrease as the period increases, whereas the FO-mode amplitudes display the expected "bell-shape" with a maximum in the middle of their instability region. Another important property of the predicted Bailey diagram is the dependence of the model period range on both the luminosity level and the chemical composition. Indeed, the period range of both the ZAHB (corresponding to the lowest luminosity level for each selected chemical composition, solid lines) and brighter (higher luminosity levels, dashed lines) models moves towards longer values as the assumed metal abundance decreases. On the other hand, at fixed metal content, brighter models show longer periods than ZAHB ones. The same trends are found when evolved models are considered (see Figure~\ref{fig-bailey-evolved}). Here the predicted period range is shifted to still longer values, even above 1 day.  However, the smaller pulsation amplitudes predicted for more metal-rich RR Lyrae allows us to disentangle any possible degeneracy between luminosity and metallicity effects.

\begin{table*}
\centering
\begin{tabular}{lcccccccc}
  \hline
Mode & N & Band$_{LSST}$ & $R^2$ & $\sigma$ & $\alpha$ & $\beta$ & $\gamma$ & $\delta$\\ \hline
  \hline
    \multicolumn{9}{c}{$Mag=\alpha+\beta{\log{P}}+\gamma[Fe/H]$} \\
F & 155 & $r$ & 0.750 & 0.128 & $0.25 \pm 0.02$ & $-1.35 \pm 0.08$ & $0.163 \pm 0.014$ & - \\
  FO & 93 & $r$ & 0.916 & 0.075 & $-0.19 \pm 0.03$ & $-1.66 \pm 0.06$ & $0.149 \pm 0.011$ & -\\ 
  FO (F slope) & 93 & $r$ & 0.650 & 0.084 & $-0.07 \pm 0.02$ & $-1.35 \pm 0.09$ & $0.156 \pm 0.012$ & -\\ 
  F+FO  & 248 & $r$ & 0.733 & 0.132 & $0.22 \pm 0.02$ & $-1.25 \pm 0.06$ & $0.168 \pm 0.012$ & - \\ 
  F & 155 & $i$ & 0.863 & 0.099 & $0.21 \pm 0.02$ & $-1.6 \pm 0.07$ & $0.163 \pm 0.011$ & -\\ 
  FO & 93 & $i$ & 0.955 & 0.059 & $-0.22 \pm 0.02$ & $-1.87 \pm 0.05$ & $0.149 \pm 0.008$ & -\\ 
  FO (F slope) & 93 & $i$ & 0.738 & 0.067 & $-0.12 \pm 0.014$ & $-1.6 \pm 0.07$ & $0.154 \pm 0.01$ & -\\ 
  F+FO  & 248 & $i$ & 0.850 & 0.105 & $0.18 \pm 0.02$ & $-1.52 \pm 0.05$ & $0.166 \pm 0.009$ & -\\ 
  F & 155 & $z$ & 0.903 & 0.087 & $0.23 \pm 0.02$ & $-1.73 \pm 0.06$ & $0.168 \pm 0.01$ & -\\ 
  FO & 93 & $z$ & 0.968 & 0.052 & $-0.21 \pm 0.02$ & $-1.99 \pm 0.04$ & $0.152 \pm 0.008$ & -\\ 
  FO (F slope) & 93 & $z$ & 0.780 & 0.061 & $-0.104 \pm 0.013$ & $-1.73 \pm 0.06$ & $0.158 \pm 0.009$ & -\\ 
  F+FO  & 248 & $z$ & 0.893 & 0.092 & $0.205 \pm 0.014$ & $-1.67 \pm 0.04$ & $0.17 \pm 0.008$ & -\\ 
  F & 155 & $y$ & 0.905 & 0.086 & $0.23 \pm 0.02$ & $-1.75 \pm 0.06$ & $0.167 \pm 0.01$ & -\\ 
  FO & 93 & $y$ & 0.967 & 0.053 & $-0.2 \pm 0.02$ & $-2 \pm 0.04$ & $0.151 \pm 0.008$ & -\\ 
  FO (F slope) & 93 & $y$ & 0.782 & 0.061 & $-0.109 \pm 0.013$ & $-1.75 \pm 0.06$ & $0.157 \pm 0.009$ & -\\ 
  F+FO  & 248 & $y$ & 0.896 & 0.092 & $0.204 \pm 0.014$ & $-1.68 \pm 0.04$ & $0.169 \pm 0.008$ & -\\ 
      \hline
  \hline
    \multicolumn{9}{c}{$W=\alpha+\beta{\log{P}}+\gamma[Fe/H]$} \\
  F & 155 & $g,u-g$ & 0.468 & 0.219 & $-3.28 \pm 0.04$ & $-1.29 \pm 0.14$ & $-0.22 \pm 0.02$ & -\\ 
  FO & 93 & $g,u-g$ & 0.593 & 0.145 & $-3.58 \pm 0.06$ & $-1.33 \pm 0.12$ & $-0.1 \pm 0.02$ & -\\ 
  FO (F slope) & 93 & $g,u-g$ & 0.219 & 0.144 & $-3.56 \pm 0.03$ & $-1.29 \pm 0.14$ & $-0.1 \pm 0.02$ & -\\ 
  F+FO & 248 & $g,u-g$ & 0.326 & 0.229 & $-3.25 \pm 0.04$ & $-0.94 \pm 0.11$ & $-0.16 \pm 0.02$ & -\\ 
  F   &         155 & $r,u-r$ &   0.869 & 0.010  &  $-1.622\pm 0.019$ & $-2.09\pm  0.07$ & $-0.059\pm   0.011$     &      -  \\     
  FO   &          93 & $r,u-r$ &   0.926 & 0.073  &  $-1.95\pm 0.03$ & $-2.07\pm  0.06$ & $-0.010\pm   0.010$     &      -   \\         
  FO (F slope) &  93 & $r,u-r$ &   0.011 & 0.073  &  $-1.960\pm 0.015$ & $-2.09\pm  0.07$ & $-0.011\pm   0.010$     &      -   \\         
  F+FO     &     248 & $r,u-r$ &   0.841 & 0.109  &  $-1.606\pm 0.017$ & $-1.89\pm  0.05$ & $-0.031\pm   0.010$     &      - \\         
  F & 155 & $r,g-r$ & 0.961 & 0.068 & $-0.489 \pm 0.013$ & $-2.63 \pm 0.04$ & $0.047 \pm 0.008$ & -\\ 
  FO & 93 & $r,g-r$ & 0.980 & 0.048 & $-0.85 \pm 0.02$ & $-2.58 \pm 0.04$ & $0.054 \pm 0.007$ & -\\ 
  FO (F slope) & 93 & $r,g-r$ & 0.399 & 0.048 & $-0.867 \pm 0.01$ & $-2.63 \pm 0.04$ & $0.053 \pm 0.007$ & -\\ 
  F+FO & 248 & $r,g-r$ & 0.967 & 0.064 & $-0.486 \pm 0.01$ & $-2.54 \pm 0.03$ & $0.053 \pm 0.006$ & -\\ 
  F & 155 & $i,g-i$ & 0.982 & 0.046 & $-0.178 \pm 0.009$ & $-2.51 \pm 0.03$ & $0.11 \pm 0.005$ & -\\ 
  FO & 93 & $i,g-i$ & 0.990 & 0.035 & $-0.565 \pm 0.013$ & $-2.56 \pm 0.03$ & $0.104 \pm 0.005$ & -\\ 
  FO (F slope) & 93 & $i,g-i$ & 0.830 & 0.035 & $-0.546 \pm 0.007$ & $-2.51 \pm 0.03$ & $0.105 \pm 0.005$ & -\\ 
  F+FO & 248 & $i,g-i$ & 0.983 & 0.046 & $-0.186 \pm 0.007$ & $-2.47 \pm 0.02$ & $0.111 \pm 0.004$ & - \\ 
  F & 155 & $z,i-z$ & 0.975 & 0.051 & $0.295 \pm 0.01$ & $-2.13 \pm 0.03$ & $0.185 \pm 0.006$ & -\\ 
  FO & 93 & $z,i-z$ & 0.990 & 0.033 & $-0.146 \pm 0.013$ & $-2.38 \pm 0.03$ & $0.164 \pm 0.005$ & -\\ 
  FO (F slope) & 93 & $z,i-z$ & 0.884 & 0.045 & $-0.052 \pm 0.009$ & $-2.13 \pm 0.03$ & $0.169 \pm 0.006$ & -\\ 
  F+FO & 248 & $z,i-z$ & 0.972 & 0.056 & $0.269 \pm 0.009$ & $-2.14 \pm 0.03$ & $0.181 \pm 0.005$ & -\\ 
  F & 155 & $y,g-y$ & 0.983 & 0.042 & $0.071 \pm 0.008$ & $-2.23 \pm 0.03$ & $0.147 \pm 0.005$ & -\\ 
  FO & 93 & $y,g-y$ & 0.990 & 0.032 & $-0.341 \pm 0.013$ & $-2.37 \pm 0.03$ & $0.134 \pm 0.005$ & -\\ 
  FO (F slope) & 93 & $y,g-y$ & 0.883 & 0.037 & $-0.288 \pm 0.008$ & $-2.23 \pm 0.03$ & $0.136 \pm 0.005$ & -\\ 
  F+FO & 248 & $y,g-y$ & 0.977 & 0.050 & $0.054 \pm 0.008$ & $-2.19 \pm 0.02$ & $0.147 \pm 0.004$ & - \\ 
      \hline
  \hline
     \multicolumn{9}{c}{$Mg=\alpha+\gamma[Fe/H]+\delta[Fe/H]^2$} \\
F+FO  & 248 & $g$ & 0.384 & 0.209 & $0.62 \pm 0.03$ & $0 \pm 0$ & $0.23 \pm 0.02$ & -\\ 
  F+FO  & 248 & $g$  & 0.396 & 0.208 & $0.67 \pm 0.03$ & $0 \pm 0$ & $0.36 \pm 0.06$ & $0.06 \pm 0.03$\\ 
   \hline
\end{tabular}
\caption{The coefficients of the fitted PLZ ($Mag = \alpha + \beta\log P + \gamma [FeH]$), PWZ ($W = \alpha + \beta\log P + \gamma [FeH]$) and absolute G magnitude versus [Fe/H] (in both the linear $M_g = \alpha  + \gamma[FeH]$ and the quadratic $M_g = \alpha  + \gamma[FeH] +\delta{[Fe/H]^2}$ form). The following quantities are listed: (column 1) the mode of the fitted sources; (column 2) the number of models used in the fit;
(column 3) the photometric band; (column 4) the coefficient of determination $R^2$; (column 5) the rms of residuals around the fitted relation; (columns 6-9) the parameters of the different fitted relations.} 
\label{tab-plzCoeffs}
\end{table*}

\begin{figure*}
   \centering
\includegraphics[width=18cm]{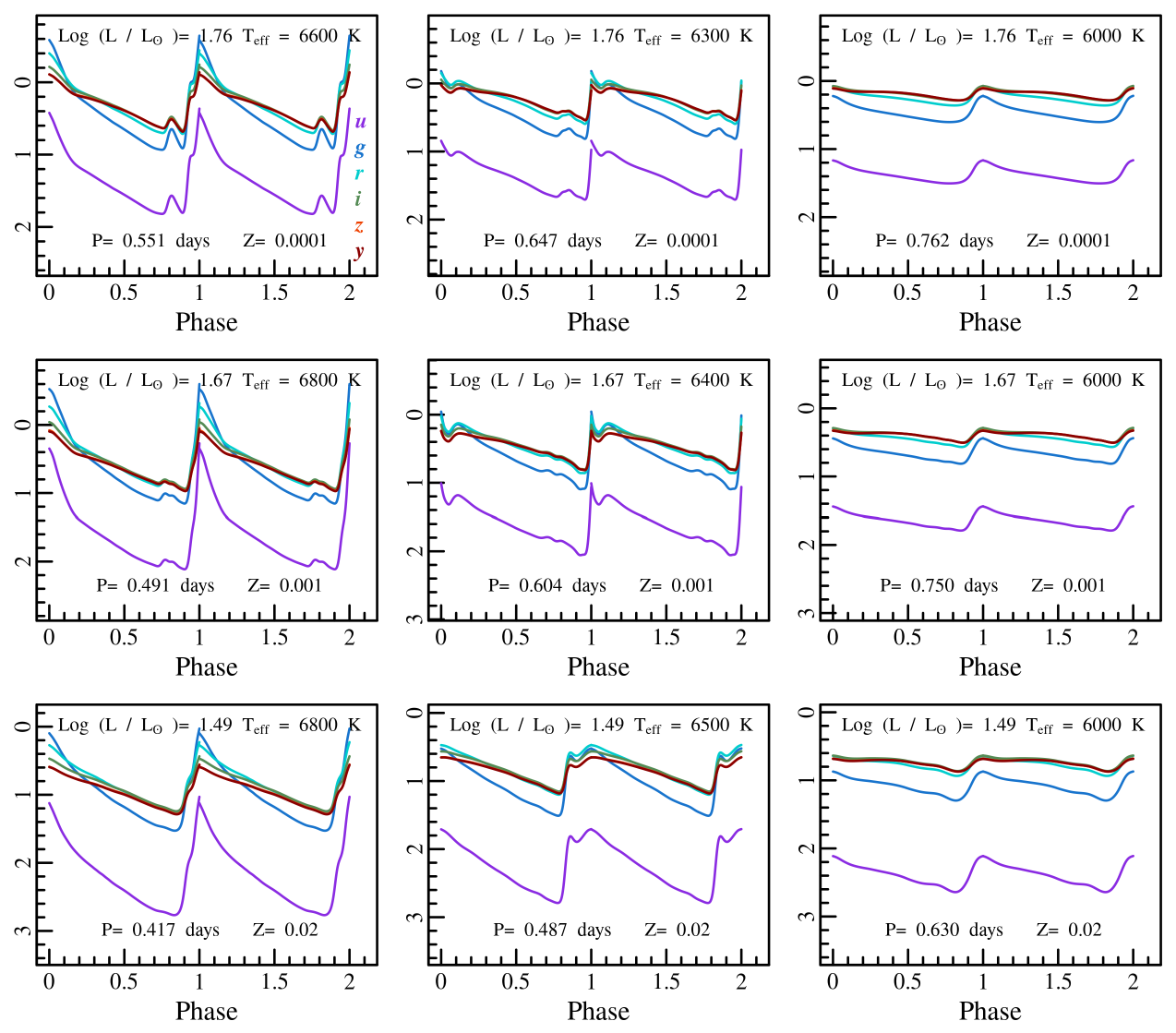}
      \caption{A subset of the derived theoretical light curves in the Rubin LSST filters the for F mode RR Lyrae. The panels in the same row show models with the same metallicity: Z = 0.0001 (top row), Z = 0.001 (middle row) and Z = 0.02 (bottom row). At fixed metallicity, the panels in different columns show models with different effective temperatures: the highest $T_{eff}$ value (left column), a middle $T_{eff}$ value (central column) and the lowest $T_{eff}$ value (right column). The pulsational period and the luminosity level are also labeled for each plotted model. The adopted bands are labelled in the top-left panel.}
         \label{fig-lc-F}
   \end{figure*}

\begin{figure*}
   \centering
\includegraphics[width=18cm]{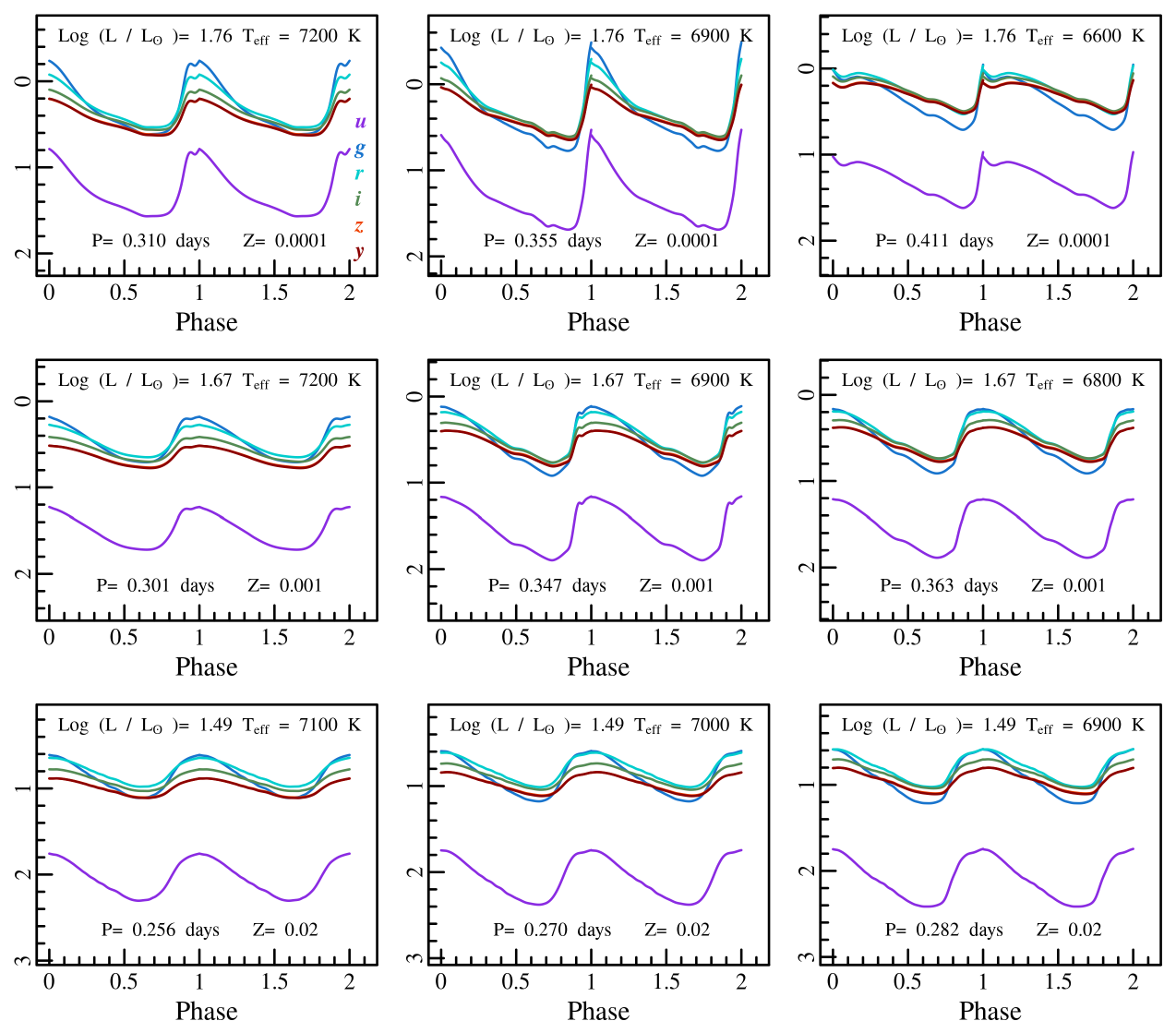}
      \caption{This figure is the same as Fig.\ref{fig-lc-F} but for the First Overtone models.}
         \label{fig-lc-FO}
   \end{figure*}

\begin{figure}
   \centering
\includegraphics[width=8.5cm]{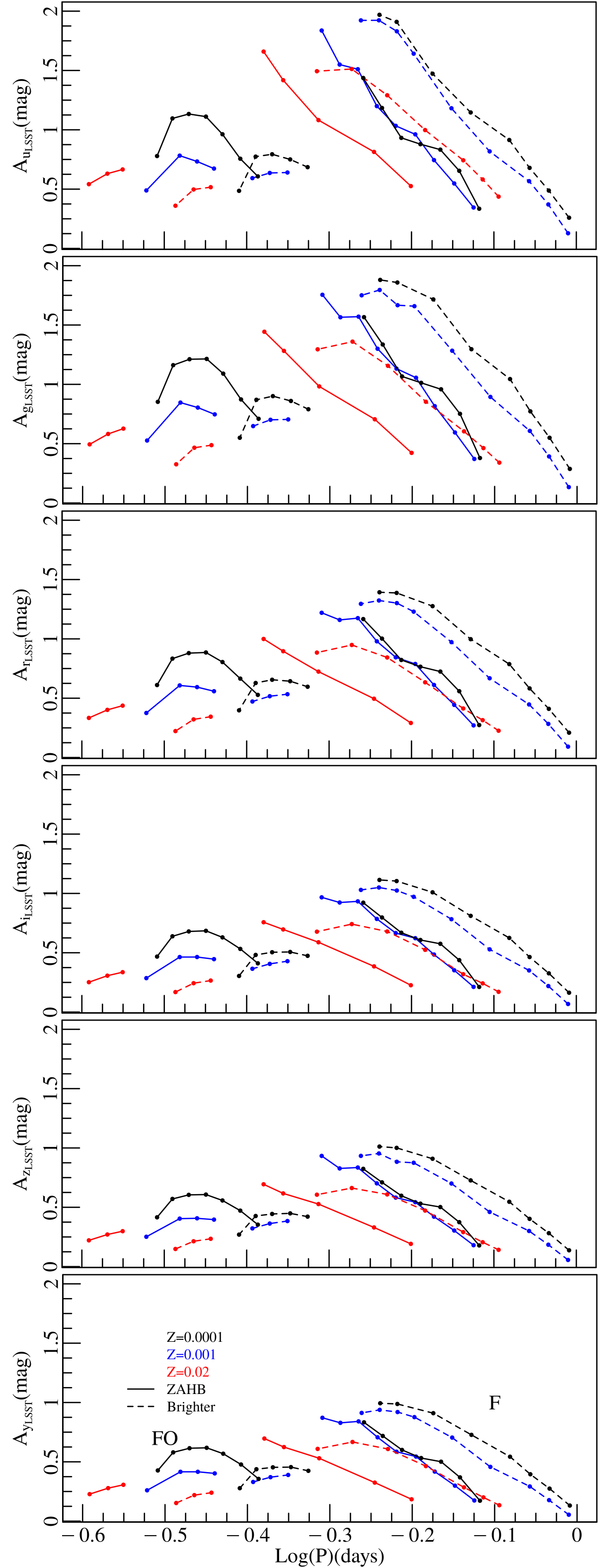}
\caption{The Bailey diagram for the ZAHB and brighter models varying the chemical composition (see labels) for both the F and the FO mode.}
\label{fig-bailey}
   \end{figure}

\begin{figure}
   \centering
\includegraphics[width=8.5cm]{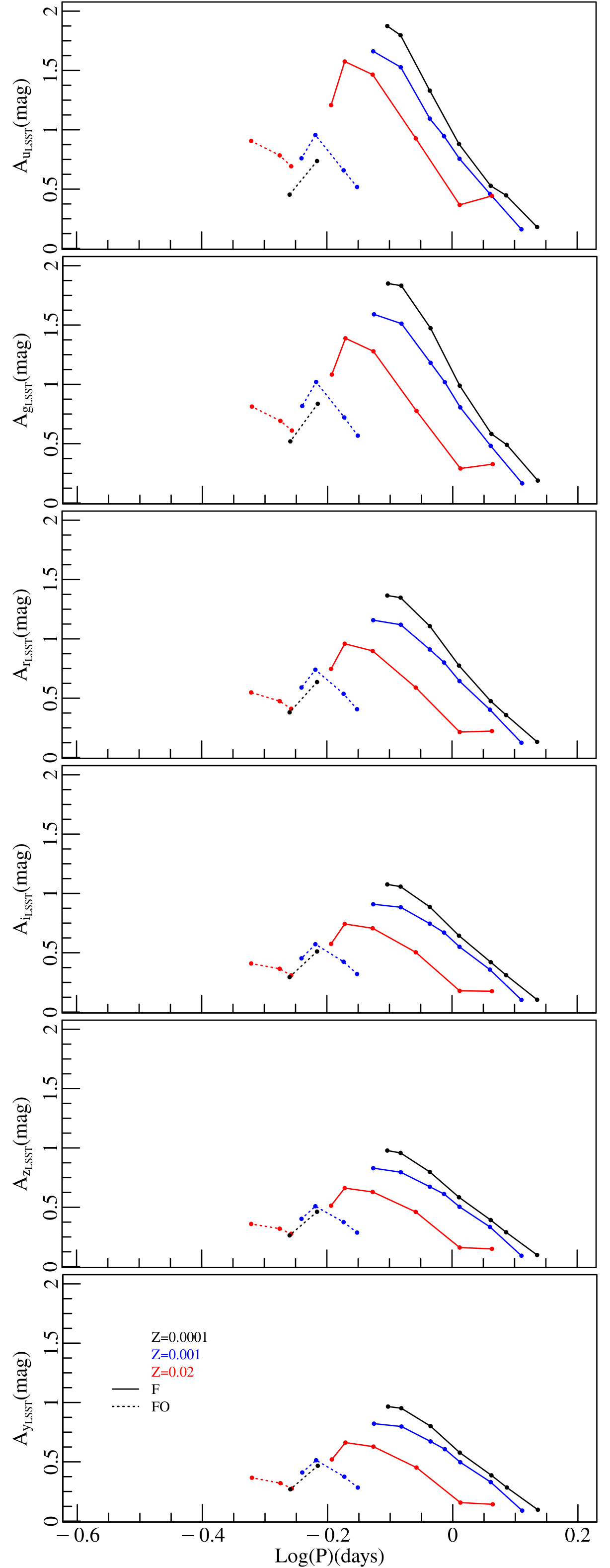}
\caption{The Bailey diagram for the Evolved models}
\label{fig-bailey-evolved}
   \end{figure}

\begin{figure*}
   \centering
\includegraphics[width=15.0cm]{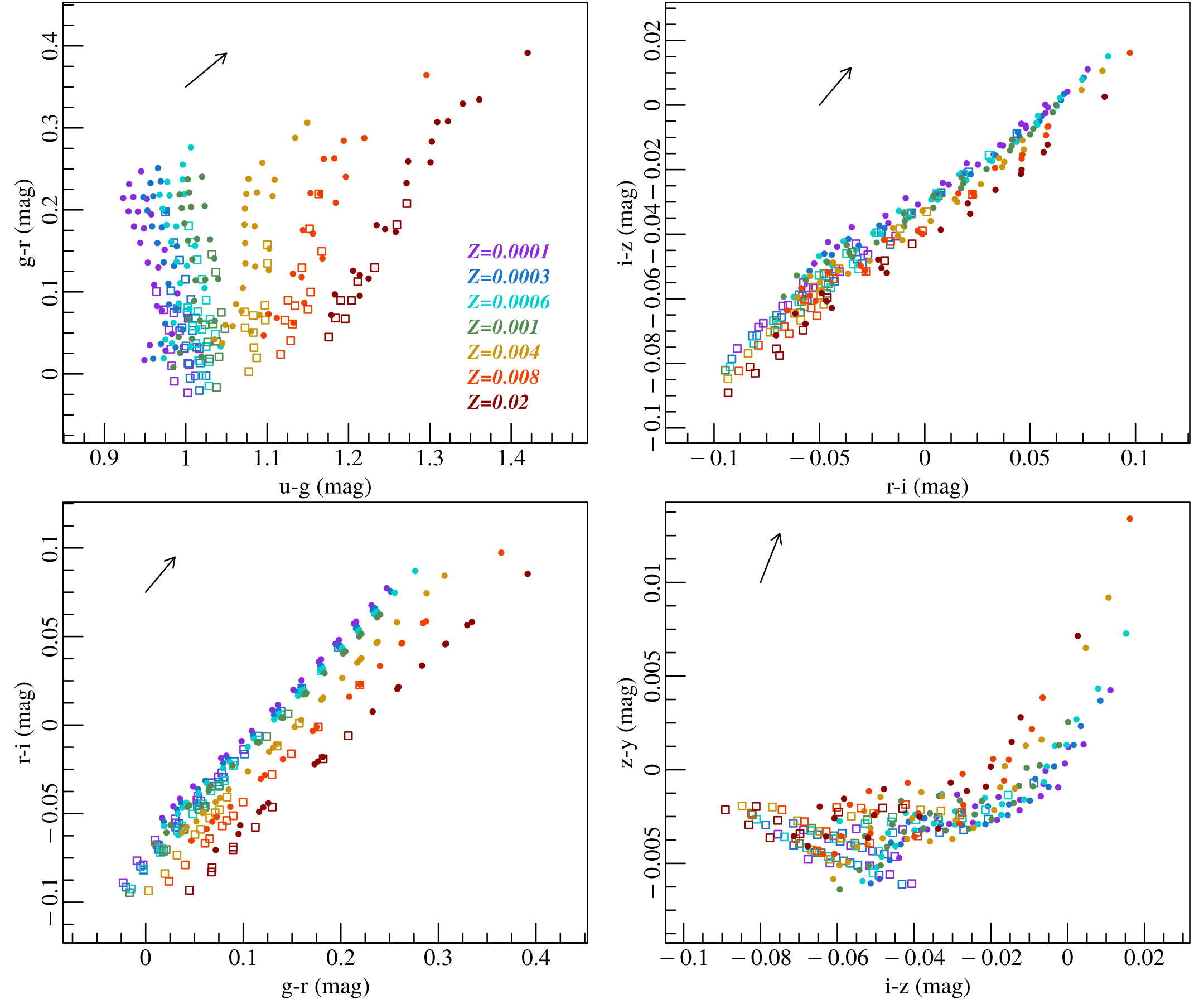}
      \caption{Theoretical color-color diagrams for different color couples: $g_{LSST}-r_{LSST}$ versus $u_{LSST}-g_{LSST}$ (top left panel), $r_{LSST}-i_{LSST}$ versus $g_{LSST}-r_{LSST}$ (bottom left panel), $i_{LSST}-z_{LSST}$ versus $r_{LSST}-i_{LSST}$ (top right panel) and $z_{LSST}-y_{LSST}$ versus $i_{LSST}-z_{LSST}$ (bottom right panel). In each panel F and FO models are plotted with filled circles and empty squares, respectively. Different colors are used to represent different Z values, as labeled in the top left panel. In each panel the black arrow indicates the reddening vector obtained by assuming the extinction law by \citet{car89}.}
         \label{fig-color-color}
   \end{figure*}

\begin{figure*}
   \centering
\includegraphics[width=15.0cm]{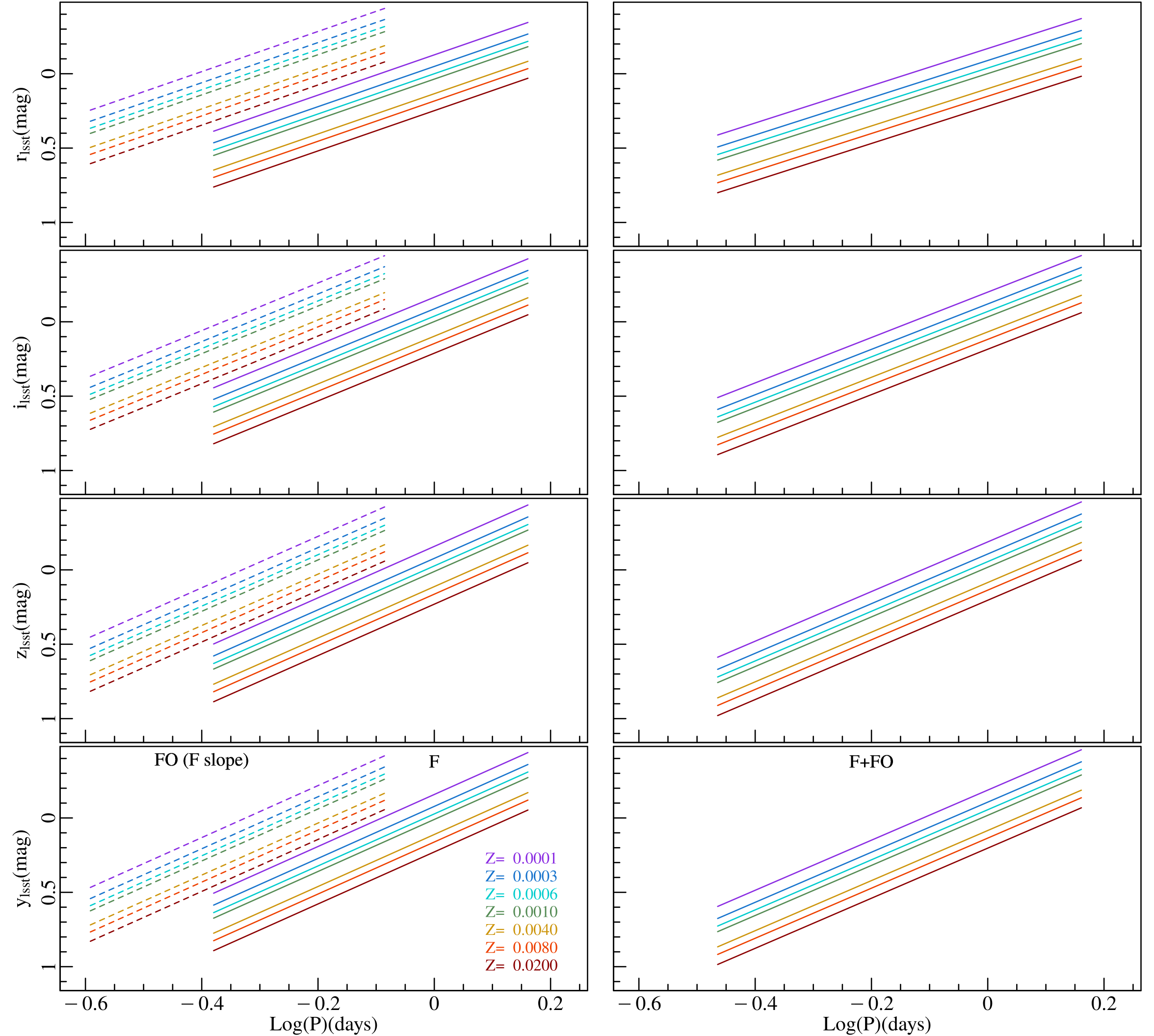}
      \caption{The predicted multi-filter PL relations for F and FO (left panels) and combined pulsators (right panels) varying the metallicity (see labels). The plotted FO mode relations have been derived with the F-mode slope, whereas in the combined relations FO-mode models have been fundamentalized.}
         \label{fig-plz}
   \end{figure*}

\begin{figure*}
   \centering
\includegraphics[width=15.0cm]{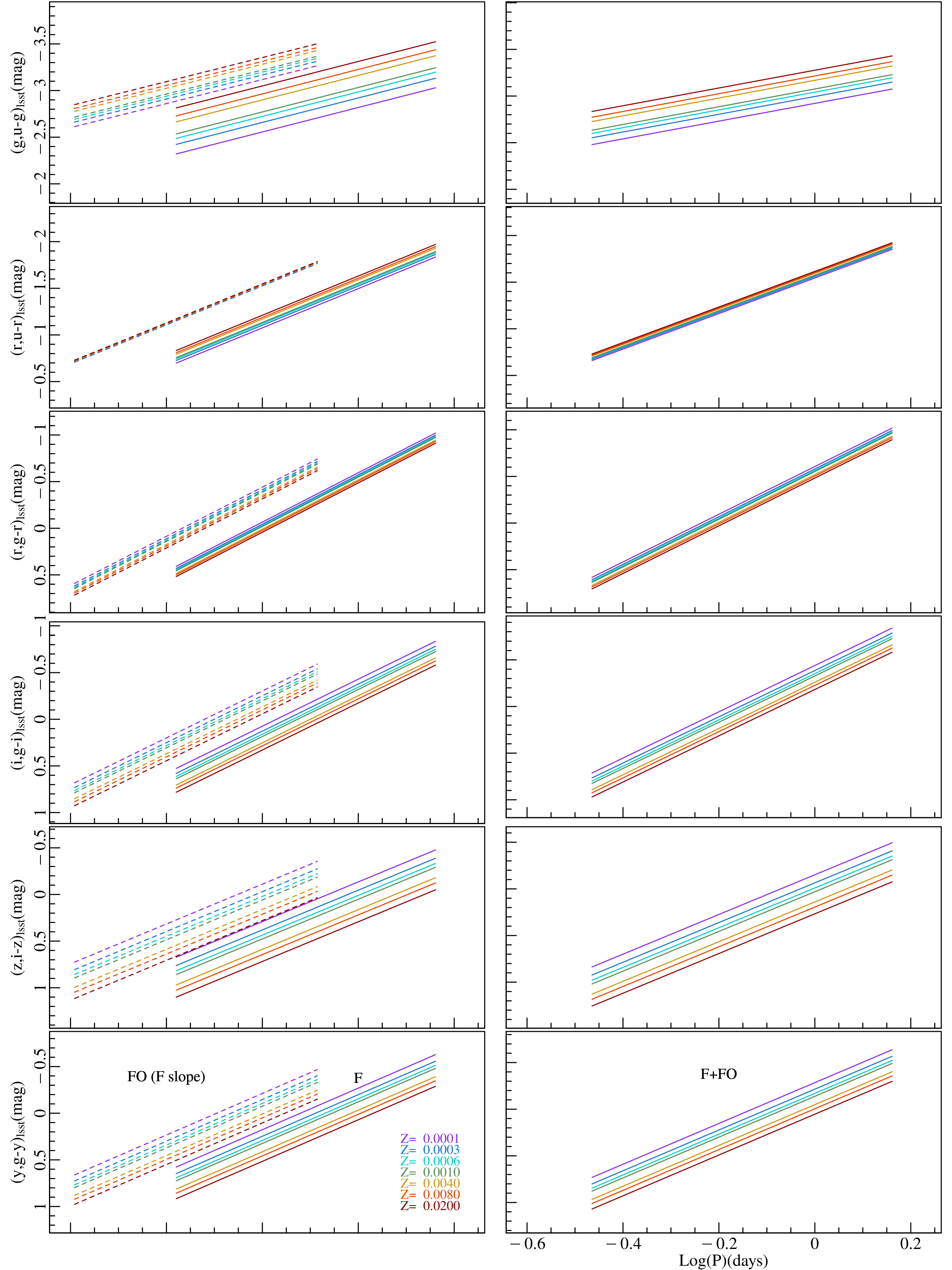}
      \caption{The same as in Figure 5 but for multi-filter PW relations.}
         \label{fig-plz-wes}
   \end{figure*}

\section{The color-color plane}\label{sec-colcol}

The derived multi-filter intensity-weighted mean magnitudes can be used to derive theoretical colors and, in turn, to investigate the behaviour of pulsation models in color-color diagrams, following similar approaches as in previous investigations in the Sloan Sky Digital Survey filters \citep[see e.g.][and references therein]{Izevic2000,Marconi06}. In Figure 7 we show the predicted distribution of both F (filled circles) and FO (open squares) RR Lyrae stars in various color-color diagrams, for the labelled Z values. In each panel the black arrow indicating the reddening vector obtained by using the extinction law by \citet{car89} is shown.
We notice that the highest sensitivity to metallicity is shown in the $g_{LSST}-r_{LSST}$ versus $u_{LSST}-g_{LSST}$ plane, but the best defined relations are obtained in the $r_{LSST}-i_{LSST}$ versus $g_{LSST}-r_{LSST}$ diagram. The latter has the advantage of producing well separated (in metallicity) linear distributions with slopes that are similar to each other and consistent with the reddening vector one. This occurrence implies that the relation between observed $r_{LSST}-i_{LSST}$ and $g_{LSST}-r_{LSST}$ is not expected to be signiﬁcantly aﬀected by uncertainties in reddening
estimates. Indeed, a systematic increase in the reddening or the
possible presence of differential reddening moves the RR Lyrae along the
same relation.  The coeﬃcients of the ﬁtted relations involving the  $r_{LSST}-i_{LSST}$ and $g_{LSST}-r_{LSST}$ colors, are listed in Table 5.
. Together with the equation containing one color as a function of the second and of the [Fe/H] values (first two lines), we fitted a linear relation expressing the metallicity content as a function of the two considered colors (last row). All the relations were obtained by considering both F- and FO-modes, but the results do not change significantly by considering only F models. As expected, the fit details contained in Table~\ref{tab-ColColZFit} suggest a strong correlation ($R^2$ close to 1) between the two considered colors and the [Fe/H] values. The relation expressing the metallicity content as a function of the two colors shows a slightly smaller correlation coefficient ($R^2=0.82$) with respect to the other two equations but appears to be an interesting tool to infer the metal abundance from $g_{LSST}, r_{LSST}, i_{LSST}$ photometry with an expected negligible dependence on uncertainties in reddening determinations (see above). 

As for the $i_{LSST}-z_{LSST}$ versus $r_{LSST}-i_{LSST}$, this relation shows the least dependence on the adopted metal content  but its slope is not significantly different from the one of the reddening vector  so that this plane is not useful for individual reddening determinations.

\begin{table*}
\centering
\begin{tabular}{ccccccc}
  \hline
Col1 & Col2 & $\alpha$ & $\beta$ & $\gamma$ & $\sigma$ & $R^2$\\ 
  \hline
    \hline
    \multicolumn{7}{c}{$Col_2=\alpha \cdot Col_1 +\beta [Fe/H] +\gamma$} \\
   $(g-r)_{LSST}$ & $(r-i)_{LSST}$ & $0.562 \pm 0.006$ & $-0.0215 \pm 0.0007$ & $-0.1089 \pm 0.0013$ & 0.008 & 0.97\\ 
   $(r-i)_{LSST}$ & $(g-r)_{LSST}$ & $1.73 \pm 0.02$ & $0.0381 \pm 0.0012$ & $0.193 \pm 0.002$ & 0.013 & 0.98 \\ 
    \hline
    \hline
    \multicolumn{7}{c}{$[Fe/H]=\alpha \cdot Col_1 +\beta \cdot Col_2 +\gamma$} \\
   $(g-r)_{LSST}$ & $(r-i)_{LSST}$ & $21.4 \pm 0.6$ & $-37 \pm 1$ & $-4.36 \pm 0.1$ & 0.313 & 0.82 \\ 
   \hline
\end{tabular}
\caption{This table contains the fitting results for color-color-[FeH] relations. Columns 1 and 2 contain the fitted colors, while the coefficients of the obtained relations are reported in columns 3-5. Finally, columns 6 and 7 contain the rms of the residuals around the fit and the coefficient of determination, respectively.} 
\label{tab-ColColZFit}
\end{table*}

\section{Metal-dependent PL and PW relations in the Rubin-LSST filters}\label{sec-pl-pw}
The obtained intensity-weighted mean magnitudes and colors can be combined with the model periods to infer multi-filter PL and PW relations. The Wesenheit magnitude combinations considered in this work are defined as follows: i) $ g_{LSST} -3.100\cdot(u_{LSST}-g_{LSST})$; ii) $ r_{LSST} -2.796\cdot(g_{LSST}-r_{LSST})$; iii) $ i_{LSST} -1.287\cdot(g_{LSST}-i_{LSST})$; iv) $ z_{LSST} -3.204\cdot(i_{LSST}-z_{LSST})$; v)  $ y_{LSST} -0.560\cdot(g_{LSST}-y_{LSST})$, with color term coefficients resulting from the extinction law by \citet{car89}, assuming $ R_V=3.1$,  while the central band wavelengths for all the LSST filters were obtained from \citet{che19} and are equal to: 0.3592 $\mu m$ ($u_{LSST}$), 0.4790 $\mu m$ ($g_{LSST}$), 0.6199 $\mu m$ ($r_{LSST}$), 0.7528 $\mu m$ ($i_{LSST}$), 0.8690 $\mu m$ ($z_{LSST}$), 0.9674 $\mu m$ ($y_{LSST}$).

As well known, only in the NIR bands RR Lyrae obey to 
tight PL relations, even if linear relations already appear around the $R/r$ wavelength \citep{2004ApJS..154..633C,2015ApJ...808...50M,2015ApJ...799..165B}. In the optical range, for a given chemical composition, their magnitude level is almost constant across the instability strip. For this reason, only the PL relations in the $r_{LSST}$, $i_{LSST}$, $z_{LSST}$ and $y_{LSST}$ bands were derived. 
To take into account the metallicity effect, we included the corresponding $[Fe/H]$ term in the linear regression and derived the first theoretical PLZ relation ($ Mag = \alpha + \beta\log(P) + \gamma [Fe/H] $) in the $r_{LSST}$, $i_{LSST}$, $z_{LSST}$ and $y_{LSST}$ bands, as reported in Table~\ref{tab-plzCoeffs}
for F, FO and combined (F $+$ fundamentalized\footnote{The fundamentalized FO periods are the periods the models would have if they pulsated in the F mode and are obtained as $\log{P_{FO}} + 0.127$ dex.} FO) model sets. We notice that, in order to reduce uncertainties related to the poorer ad not uniform number of FO models especially as the luminosity increases, the FO mode relations have also been obtained fixing the slope of the F relations. These fixed slope relations are plotted in Figure~\ref{fig-plz} together with the F mode and the combined ones.
We notice that the metallicity effect on the relation zero points is not negligible with more metal rich RR Lyrae models providing fainter magnitudes at fixed period in all the investigated bands.
As well known, only the zero point $\alpha$ is affected by this transformation.
As for the PW relations, following a similar approach as in \citet{2021MNRAS.500.5009M}, where multi-fiter PWZ relations in the Gaia bands were derived, in Table~\ref{tab-plzCoeffs} we also report the coefficients of the obtained PWZ relations in the Rubin-LSST filters. The obtained relations are plotted in Figure 6.  We notice that the metallicity effect depends on the band combinations and is minimum for the $g_{LSST}$,$g_{LSST}-r_{LSST}$ and $r_{LSST}$,$u_{LSST}-r_{LSST}$ cases. For the latter combination the FO relation appears to be independent of metallicity.
These results are in agreement with the almost negligible metallicity dependence found for the $V,B-V$ PW relation in \citet{2015ApJ...808...50M} and makes the $g_{LSST}$,$g_{LSST}-r_{LSST}$ and $r_{LSST}$,$u_{LSST}-r_{LSST}$ filter combinations particularly useful to use RR Lyrae that will be observed by the Rubin-LSST as standard candles, in spite of their metallicity uncertainties. We also notice an interesting inversion of the metallicity effect in the PW relation involving the $u_{LSST}$ band.

\section{The $\lowercase{g}_{LSST}-[Fe/H]$ relation}\label{sec-g-feh}

The inferred intensity weighted mean $g_{LSST}$ magnitudes also allow us to derive the first predicted $g_{LSST}-[Fe/H]$ relation. The coefficients of the traditional linear form and of the hypothesized quadratic one are reported in the last section of Table 3 combining the F and FO model sets.
Indeed, some previous works suggested that this relation could not be linear over the whole observed metallicity range of RR Lyrae stars \citep[see e.g.][and references therein]{Caputo2000,2004ApJ...612.1092D} even if the linear form is the most used in the literature.
Indeed according to the relations reported in Table 3, the linear form seems a reasonably good approximation of the model behaviour, as the quadratic term is smaller than the linear one (about 15 \%).
We also notice that the metallicity coefficient in the linear relation is, as expected, consistent within the errors with the one predicted in the V band, e.g. in \citet{2018ApJ...864L..13M} but smaller than the empirical determination by \citet{2018MNRAS.481.1195M}.
However, we recall that in spite of the advantage of directly correlating the absolute optical magnitude to a measured metal content, without the contribution of color terms, this relation suffers from a number of drawbacks, e.g. the possible systematic effect produced by the unknown evolutionary status of observed RR Lyrae stars or the adopted metallicity scale  and $\alpha$ element enhancement.

\section{Final Remarks}

With a single visit depth of g $\sim$ 24.91 mag and about 1000 repeated observations over a 10-year period the Rubin LSST will provide  an opportunity to measure the distribution of RR Lyrae within different galactic and extragalactic environments. In this context, similarly to the PLZ and PWZ relations derived in the Johnson-Cousins and Gaia bands \citep[see][]{2015ApJ...807..127M,2021MNRAS.500.5009M}, the predicted Rubin-LSST PLZ relations represent a powerful tool to derive mean and individual distances of Galactic and extragalactic RR Lyrae stars, once periods and accurate spectroscopic metal abundances are available. On the other hand, for RR Lyrae at the same distance (e.g. belonging to the same stellar cluster or galaxy), the obtained relations can be used to derive the metallicity distribution of the investigated sample \citep[see e.g.][]{2016AJ....152..170B,2016MNRAS.461L..41M,2021MNRAS.508.1064M}. Finally, once accurate estimates of individual distances are available, the derived PLZ relations are excellent tools to provide the actual metallicity values of Galactic RR Lyrae from the observed periods and Rubin-LSST mean magnitudes.
An interesting implication of the inferred theoretical relations is the predicted systematic effect due to possible metallicity differences among the investigated RR Lyrae samples. In particular, if a mean metallicity is assumed for RR Lyrae stars in a given stellar system, the individual distance moduli obtained from the application of the PLZ relations in Table 1, will be underestimated or overestimated by and amount $\Delta mag$ for any given $\Delta[Fe/H]$ between the true metal abundance and the assumed mean one, with ${\Delta mag}/{\Delta[Fe/H]}\sim 0.2$ mag/dex in the case of F-mode and $\sim$ 0.1 mag/dex in the case of FO-mode RR Lyrae stars. On this basis, in order to best exploit the predictive power of the inferred relations accurate spectroscopic individual abundances are needed \citep{2019MNRAS.487.5463C}. Moreover, the contribution of $\alpha$ element enhancement in the metal poor stellar populations which RR Lyrae belong to, should be taken into account, as the inclusion of $\alpha$ elements contributions simulates the adoption of a higher global metallicity \citep[see e.g.][]{1993ApJ...414..580S} in the adopted pulsation models. Moreover the adopted bolometric corrections have been found to depend on the adopted $\alpha$ enhancement \citep[see e.g.][]{Cassisi2004}. As for the helium content, we have shown in \citet{2018ApJ...864L..13M} that the main effect of an increase in the helium abundance is a brighter ZAHB at fixed metallicity and mass, with consequent longer periods and non negligible effects on the coefficients of predicted relations. However, especially towards the longer wavelengths, the final effect in the inferred individual distances is expected to be similar to the standard deviation of the adopted e.g. PLZ relations \citep[see][for details]{2018ApJ...864L..13M}.


\acknowledgments
\textbf{We thank the anonymous Referee for her/his comments that contributed to improve the content of the manuscript.}

\bibliography{mar}{}

\begin{thebibliography}{}
\expandafter\ifx\csname natexlab\endcsname\relax\def\natexlab#1{#1}\fi
\providecommand{\url}[1]{\href{#1}{#1}}
\providecommand{\dodoi}[1]{doi:~\href{http://doi.org/#1}{\nolinkurl{#1}}}
\providecommand{\doeprint}[1]{\href{http://ascl.net/#1}{\nolinkurl{http://ascl.net/#1}}}
\providecommand{\doarXiv}[1]{\href{https://arxiv.org/abs/#1}{\nolinkurl{https://arxiv.org/abs/#1}}}

\bibitem[{{Bono} {et~al.}(2001){Bono}, {Caputo}, {Castellani}, {Marconi}, \&
  {Storm}}]{2001MNRAS.326.1183B}
{Bono}, G., {Caputo}, F., {Castellani}, V., {Marconi}, M., \& {Storm}, J. 2001,
  \mnras, 326, 1183, \dodoi{10.1046/j.1365-8711.2001.04655.x}

\bibitem[{{Bono} {et~al.}(2003){Bono}, {Caputo}, {Castellani}, {Marconi},
  {Storm}, \& {Degl'Innocenti}}]{2003MNRAS.344.1097B}
{Bono}, G., {Caputo}, F., {Castellani}, V., {et~al.} 2003, \mnras, 344, 1097,
  \dodoi{10.1046/j.1365-8711.2003.06878.x}

\bibitem[{{Braga} {et~al.}(2015){Braga}, {Dall'Ora}, {Bono}, {Stetson},
  {Ferraro}, {Iannicola}, {Marengo}, {Neeley}, {Persson}, {Buonanno},
  {Coppola}, {Freedman}, {Madore}, {Marconi}, {Matsunaga}, {Monson}, {Rich},
  {Scowcroft}, \& {Seibert}}]{2015ApJ...799..165B}
{Braga}, V.~F., {Dall'Ora}, M., {Bono}, G., {et~al.} 2015, \apj, 799, 165,
  \dodoi{10.1088/0004-637X/799/2/165}

\bibitem[{{Braga} {et~al.}(2016){Braga}, {Stetson}, {Bono}, {Dall'Ora},
  {Ferraro}, {Fiorentino}, {Freyhammer}, {Iannicola}, {Marengo}, {Neeley},
  {Valenti}, {Buonanno}, {Calamida}, {Castellani}, {da Silva},
  {Degl'Innocenti}, {Di Cecco}, {Fabrizio}, {Freedman}, {Giuffrida}, {Lub},
  {Madore}, {Marconi}, {Marinoni}, {Matsunaga}, {Monelli}, {Persson},
  {Piersimoni}, {Pietrinferni}, {Prada-Moroni}, {Pulone}, {Stellingwerf},
  {Tognelli}, \& {Walker}}]{2016AJ....152..170B}
{Braga}, V.~F., {Stetson}, P.~B., {Bono}, G., {et~al.} 2016, \aj, 152, 170,
  \dodoi{10.3847/0004-6256/152/6/170}

\bibitem[{{Caputo} {et~al.}(2000){Caputo}, {Castellani}, {Marconi}, \&
  {Ripepi}}]{Caputo2000}
{Caputo}, F., {Castellani}, V., {Marconi}, M., \& {Ripepi}, V. 2000, \mnras,
  316, 819, \dodoi{10.1046/j.1365-8711.2000.03591.x}

\bibitem[{{Cardelli} {et~al.}(1989){Cardelli}, {Clayton}, \& {Mathis}}]{car89}
{Cardelli}, J.~A., {Clayton}, G.~C., \& {Mathis}, J.~S. 1989, \apj, 345, 245,
  \dodoi{10.1086/167900}

\bibitem[{{Cassisi} {et~al.}(2021){Cassisi}, {Potekhin}, {Salaris}, \&
  {Pietrinferni}}]{Cassisi2021}
{Cassisi}, S., {Potekhin}, A.~Y., {Salaris}, M., \& {Pietrinferni}, A. 2021,
  \aap, 654, A149, \dodoi{10.1051/0004-6361/202141425}

\bibitem[{{Cassisi} {et~al.}(2004){Cassisi}, {Salaris}, {Castelli}, \&
  {Pietrinferni}}]{Cassisi2004}
{Cassisi}, S., {Salaris}, M., {Castelli}, F., \& {Pietrinferni}, A. 2004, \apj,
  616, 498, \dodoi{10.1086/424907}

\bibitem[{{Catelan} {et~al.}(2004){Catelan}, {Pritzl}, \&
  {Smith}}]{2004ApJS..154..633C}
{Catelan}, M., {Pritzl}, B.~J., \& {Smith}, H.~A. 2004, \apjs, 154, 633,
  \dodoi{10.1086/422916}

\bibitem[{{Chen} {et~al.}(2019){Chen}, {Girardi}, {Fu}, {Bressan}, {Aringer},
  {Dal Tio}, {Pastorelli}, {Marigo}, {Costa}, \& {Zhang}}]{che19}
{Chen}, Y., {Girardi}, L., {Fu}, X., {et~al.} 2019, \aap, 632, A105,
  \dodoi{10.1051/0004-6361/201936612}

\bibitem[{{Coppola} {et~al.}(2011){Coppola}, {Dall'Ora}, {Ripepi}, {Marconi},
  {Musella}, {Bono}, {Piersimoni}, {Stetson}, \& {Storm}}]{2011MNRAS.416.1056C}
{Coppola}, G., {Dall'Ora}, M., {Ripepi}, V., {et~al.} 2011, \mnras, 416, 1056,
  \dodoi{10.1111/j.1365-2966.2011.19102.x}

\bibitem[{{Crestani} {et~al.}(2019){Crestani}, {Alves-Brito}, {Bono}, {Puls},
  \& {Alonso-Garc{\'\i}a}}]{2019MNRAS.487.5463C}
{Crestani}, J., {Alves-Brito}, A., {Bono}, G., {Puls}, A.~A., \&
  {Alonso-Garc{\'\i}a}, J. 2019, \mnras, 487, 5463,
  \dodoi{10.1093/mnras/stz1674}

\bibitem[{{Dall'Ora} {et~al.}(2006){Dall'Ora}, {Bono}, {Storm}, {Caputo},
  {Andreuzzi}, {Marconi}, {Monelli}, {Ripepi}, {Stetson}, \&
  {Testa}}]{2006MmSAI..77..214D}
{Dall'Ora}, M., {Bono}, G., {Storm}, J., {et~al.} 2006, \memsai, 77, 214.
\newblock \doarXiv{astro-ph/0601237}

\bibitem[{{Di Criscienzo} {et~al.}(2004){Di Criscienzo}, {Marconi}, \&
  {Caputo}}]{2004ApJ...612.1092D}
{Di Criscienzo}, M., {Marconi}, M., \& {Caputo}, F. 2004, \apj, 612, 1092,
  \dodoi{10.1086/422742}

\bibitem[{{Ivezi{\'c}} {et~al.}(2000){Ivezi{\'c}}, {Goldston}, {Finlator},
  {Knapp}, {Yanny}, {McKay}, {Amrose}, {Krisciunas}, {Willman}, {Anderson},
  {Schaber}, {Erb}, {Logan}, {Stubbs}, {Chen}, {Neilsen}, {Uomoto}, {Pier},
  {Fan}, {Gunn}, {Lupton}, {Rockosi}, {Schlegel}, {Strauss}, {Annis},
  {Brinkmann}, {Csabai}, {Doi}, {Fukugita}, {Hennessy}, {Hindsley}, {Margon},
  {Munn}, {Newberg}, {Schneider}, {Smith}, {Szokoly}, {Thakar}, {Vogeley},
  {Waddell}, {Yasuda}, {York}, \& {SDSS Collaboration}}]{Izevic2000}
{Ivezi{\'c}}, {\v{Z}}., {Goldston}, J., {Finlator}, K., {et~al.} 2000, \aj,
  120, 963, \dodoi{10.1086/301455}

\bibitem[{{Longmore} {et~al.}(1990){Longmore}, {Dixon}, {Skillen}, {Jameson},
  \& {Fernley}}]{Longmore90}
{Longmore}, A.~J., {Dixon}, R., {Skillen}, I., {Jameson}, R.~F., \& {Fernley},
  J.~A. 1990, \mnras, 247, 684

\bibitem[{{Longmore} {et~al.}(1986){Longmore}, {Fernley}, \&
  {Jameson}}]{Longmore86}
{Longmore}, A.~J., {Fernley}, J.~A., \& {Jameson}, R.~F. 1986, \mnras, 220,
  279, \dodoi{10.1093/mnras/220.2.279}

\bibitem[{{Marconi} {et~al.}(2018){Marconi}, {Bono}, {Pietrinferni}, {Braga},
  {Castellani}, \& {Stellingwerf}}]{2018ApJ...864L..13M}
{Marconi}, M., {Bono}, G., {Pietrinferni}, A., {et~al.} 2018, \apjl, 864, L13,
  \dodoi{10.3847/2041-8213/aada17}

\bibitem[{{Marconi} {et~al.}(2006){Marconi}, {Cignoni}, {Di Criscienzo},
  {Ripepi}, {Castelli}, {Musella}, \& {Ruoppo}}]{Marconi06}
{Marconi}, M., {Cignoni}, M., {Di Criscienzo}, M., {et~al.} 2006, \mnras, 371,
  1503, \dodoi{10.1111/j.1365-2966.2006.10787.x}

\bibitem[{{Marconi} {et~al.}(2021){Marconi}, {Molinaro}, {Ripepi}, {Leccia},
  {Musella}, {De Somma}, {Gatto}, \& {Moretti}}]{2021MNRAS.500.5009M}
{Marconi}, M., {Molinaro}, R., {Ripepi}, V., {et~al.} 2021, \mnras, 500, 5009,
  \dodoi{10.1093/mnras/staa3558}

\bibitem[{{Marconi} {et~al.}(2015){Marconi}, {Coppola}, {Bono}, {Braga},
  {Pietrinferni}, {Buonanno}, {Castellani}, {Musella}, {Ripepi}, \&
  {Stellingwerf}}]{2015ApJ...808...50M}
{Marconi}, M., {Coppola}, G., {Bono}, G., {et~al.} 2015, \apj, 808, 50,
  \dodoi{10.1088/0004-637X/808/1/50}

\bibitem[{{Mart{\'\i}nez-V{\'a}zquez}
  {et~al.}(2016){Mart{\'\i}nez-V{\'a}zquez}, {Monelli}, {Gallart}, {Bono},
  {Bernard}, {Stetson}, {Ferraro}, {Walker}, {Dall'Ora}, {Fiorentino}, \&
  {Iannicola}}]{2016MNRAS.461L..41M}
{Mart{\'\i}nez-V{\'a}zquez}, C.~E., {Monelli}, M., {Gallart}, C., {et~al.}
  2016, \mnras, 461, L41, \dodoi{10.1093/mnrasl/slw093}

\bibitem[{{Mart{\'\i}nez-V{\'a}zquez}
  {et~al.}(2021){Mart{\'\i}nez-V{\'a}zquez}, {Monelli}, {Cassisi}, {Taibi},
  {Gallart}, {Vivas}, {Walker}, {Mart{\'\i}n-Ravelo}, {Zenteno}, {Battaglia},
  {Bono}, {Calamida}, {Carollo}, {Cicu{\'e}ndez}, {Fiorentino}, {Marconi},
  {Salvadori}, {Balbinot}, {Bernard}, {Dall'Ora}, \&
  {Stetson}}]{2021MNRAS.508.1064M}
{Mart{\'\i}nez-V{\'a}zquez}, C.~E., {Monelli}, M., {Cassisi}, S., {et~al.}
  2021, \mnras, 508, 1064, \dodoi{10.1093/mnras/stab2493}

\bibitem[{{Muraveva} {et~al.}(2018){Muraveva}, {Delgado}, {Clementini},
  {Sarro}, \& {Garofalo}}]{2018MNRAS.481.1195M}
{Muraveva}, T., {Delgado}, H.~E., {Clementini}, G., {Sarro}, L.~M., \&
  {Garofalo}, A. 2018, \mnras, 481, 1195, \dodoi{10.1093/mnras/sty2241}

\bibitem[{{Muraveva} {et~al.}(2015){Muraveva}, {Palmer}, {Clementini}, {Luri},
  {Cioni}, {Moretti}, {Marconi}, {Ripepi}, \& {Rubele}}]{2015ApJ...807..127M}
{Muraveva}, T., {Palmer}, M., {Clementini}, G., {et~al.} 2015, \apj, 807, 127,
  \dodoi{10.1088/0004-637X/807/2/127}

\bibitem[{{Pietrinferni} {et~al.}(2013){Pietrinferni}, {Cassisi}, {Salaris}, \&
  {Hidalgo}}]{2013AA}
{Pietrinferni}, A., {Cassisi}, S., {Salaris}, M., \& {Hidalgo}, S. 2013, \aap,
  558, A46, \dodoi{10.1051/0004-6361/201321950}

\bibitem[{{Salaris} {et~al.}(1993){Salaris}, {Chieffi}, \&
  {Straniero}}]{1993ApJ...414..580S}
{Salaris}, M., {Chieffi}, A., \& {Straniero}, O. 1993, \apj, 414, 580,
  \dodoi{10.1086/173105}

\bibitem[{{Sesar} {et~al.}(2017){Sesar}, {Fouesneau}, {Price-Whelan},
  {Bailer-Jones}, {Gould}, \& {Rix}}]{2017ApJ...838..107S}
{Sesar}, B., {Fouesneau}, M., {Price-Whelan}, A.~M., {et~al.} 2017, \apj, 838,
  107, \dodoi{10.3847/1538-4357/aa643b}

\bibitem[{{Sollima} {et~al.}(2006){Sollima}, {Cacciari}, \&
  {Valenti}}]{2006MNRAS.372.1675S}
{Sollima}, A., {Cacciari}, C., \& {Valenti}, E. 2006, \mnras, 372, 1675,
  \dodoi{10.1111/j.1365-2966.2006.10962.x}

\bibitem[{{Valle} {et~al.}(2013){Valle}, {Dell'Omodarme}, {Prada Moroni}, \&
  {Degl'Innocenti}}]{Valle2013}
{Valle}, G., {Dell'Omodarme}, M., {Prada Moroni}, P.~G., \& {Degl'Innocenti},
  S. 2013, \aap, 549, A50, \dodoi{10.1051/0004-6361/201220069}

\end{thebibliography}

\bibliographystyle{aasjournal}

\end{document}